\documentclass[aps,prl,reprint,twocolumn,amsmath,amssymb,citeautoscript,floatfix]{revtex4-1}
\usepackage{graphicx}% Include figure files
\usepackage{dcolumn}% Align table columns on decimal point
\usepackage{bm}% bold math
\usepackage{latexsym,epsfig}
\usepackage{graphicx}
\usepackage{verbatim}
\usepackage{comment}
\usepackage{amsmath}
\usepackage{amssymb}
\usepackage{physics}
\usepackage{stmaryrd}
\usepackage{color}
\usepackage{epstopdf}
\usepackage{grffile}
\usepackage{lipsum}
\usepackage{enumitem}
\usepackage{dsfont}
\DeclareGraphicsExtensions{.eps}
\newcommand{\beq}{\begin{equation}}
\newcommand{\eeq}{\end{equation}}
\newcommand{\bea}{\begin{eqnarray}}
\newcommand{\eea}{\end{eqnarray}}
\newcommand{\ben}{\begin{eqnarray*}}
\newcommand{\een}{\end{eqnarray*}}
\newcommand{\bfig}{\begin{figure}}
\newcommand{\efig}{\end{figure}}

\usepackage{hyperref}
\hypersetup{
    colorlinks=true,      
    urlcolor=blue,
    citecolor=blue,
    linkcolor=blue
}

\begin{document}
\title{Disorder driven Thouless charge pump in a quasiperiodic chain}

 \author{Ashirbad Padhan$^{1,2}$ and Tapan Mishra$^{1,2,*}$}
%\affiliation{$^{1}$Department of Physics, Indian Institute of Technology, Guwahati, Assam - 781039, India}

\affiliation{$^1$ School of Physical Sciences, National Institute of Science Education and Research, Jatni 752050, India}

\affiliation{$^2$ Homi Bhabha National Institute, Training School Complex, Anushaktinagar, Mumbai 400094, India}

\date{\today}
 \email{mishratapan@gmail.com}

\begin{abstract}
Thouless charge pump enables a quantized transport of charge through an adiabatic evolution of the Hamiltonian exhibiting topological phase. While this charge pumping is known to be robust against the presence of weak disorder in the system, it often breaks down with the increase in disorder strength. In this work, however, we show that in a one dimensional Su-Schrieffer-Heeger lattice, a unit cell-wise staggered quasiperiodic disorder favors a quantized charge pump. Moreover, we show that such quantized Thouless charge pump is achieved by following the standard single cycle pumping protocol which usually leads to a breakdown of charge pump in other known models. This unusual property is found to be due to an emergence of a trivial gapped phase from a topological phase as the quasiperiodic disorder is tuned. This emergent gapped to gapped transition also allows us to propose a non-standard pumping scheme where a modulated disorder favors a quantized Thouless charge pump.

\end{abstract}

\maketitle
 
{\em Introduction.-} Thouless charge pumps (TCP) are the quantum analog of the Archimedes screw pump which facilitates particle transport in a one dimensional lattice through adiabatic and periodic modulation of some system parameters of the system and not by any external gradient or field~\cite{thouless, QNiu_1984}. In the case of topological phase transitions through gap closing singularities, the TCPs  are known to be quantized around a cycle encircling the gap closing point and are related to the topological invariant known as the Chern number defined in the context of quantum Hall physics~\cite{Klitzing, Thouless1982, KOHMOTO1985343, Citro_2023}.  Due to the versatility of such phenomenon of charge transport, the TCP has attracted a great deal of attention in recent years and has been observed in various artificial systems to characterize topological phases and phase transitions~\cite{Nakajima_2016, Lohse_2015, Spielman, bloch_spinpump, Sylvain, Esslinger_expt, Wang_expt, Zilberberg_expt, J_rgensen_2021, J_rgensen_2023, Cheng_2022, Liu_expt, Ke_2016, Cheng_expt, Grinberg2020, tao2023interactioninduced, viebahn2023interactioninduced, Hatsugai_2016, Wang, Chaohong_2020, Mondal2020, suman_v1v2, Kuno_int, Nakagawa2018, bertok_pump, mondal_phonon, meden, Hayward2018, mondal_sshhubbard, Barbiero, Citro_2023, Cooper_rev, Ozawa_rev}. However, the presence of disorder plays a crucial role in stabilizing the TCPs. While the quantization of charge transport remains robust against weak disorder or as long as the bulk gap remains finite, strong disorder leads to a breakdown of the TCP due to the closing of the bulk gap and vanishing of the topological phase~\cite{hayward, Takahashi,tatp, floquet1d, HOTI, Cerjan2020}.

Exceptions have been found in the form of topological Anderson insulator~\cite{tai, Groth_tai, Meier_2018, tai_Jian, Altland_tai, tai3d} in recent years where a disorder in the hopping term is found to promote a topological phase starting from a clean trivial phase leading to a quantised TCP~\cite{tatp}. On the other hand, a stable TCP has been established in a chain with onsite quasiperiodic disorder by following an indirect approach of multi-cycle pumping protocol~\cite{Takahashi, tatp}. While for the case with hopping disorder, the quantized pumping is attributed to the preserved chiral symmetry in the entire process that leads to the onset of the topological nature, for the case of onsite disorder, the finite pumping is achieved by encircling the gap closing point through two different pumping paths~\cite{Takahashi,tatp}. In the latter case, the particular choice of two pumping protocols together contribute to a finite charge pump for some intermediate values of disorder although for each individual pumping protocols the pumping clearly breaks down. At this point the question still remains open whether a stable charge pump is possible in lattices with onsite disorder or not. 

\begin{figure}[!t]
\begin{center}
\includegraphics[width=1\columnwidth]{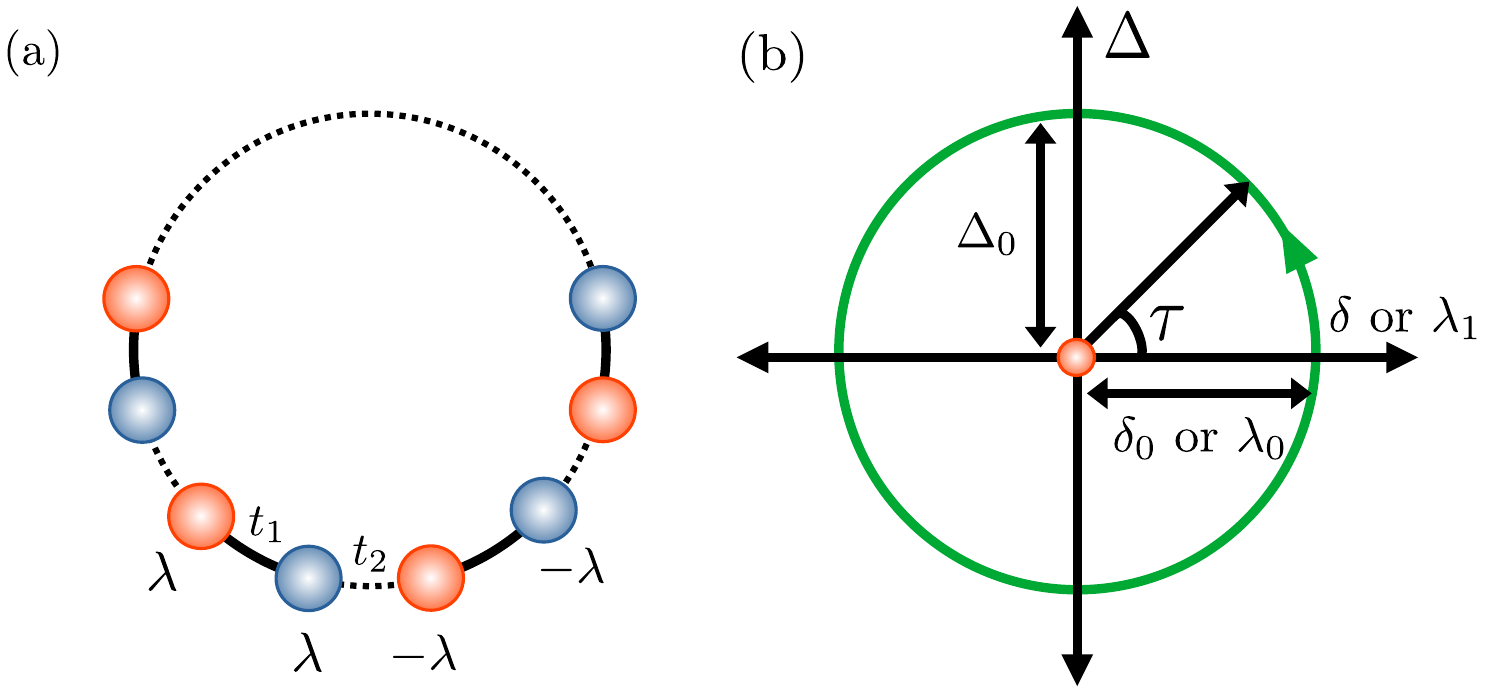}
 \end{center}
\caption{(a) The figure depicts the model shown in Eq.~\ref{eq:ham_topdis_ssh} in terms of intra-cell hopping strength $t_1$, inter-cell hopping strength $t_2$ and disorder strength $\lambda$. The red (blue) ball represents the $A$ ($B$) sublattice site of a unit cell. (b) Two types of pumping schemes are shown in either $\delta-\Delta$ plane or $\lambda_1-\Delta$ plane. The green circle with an arrow represents the pumping path and the red dot marks a gapless critical point. (See text for details.)}  
\label{fig:figure0}
\end{figure}

In this work, we propose a one dimensional model with onsite quasiperiodic disorder where the TCP does not break down, rather a quantized Thouless charge pump due to disorder is possible by following the conventional single cycle pumping protocol. By considering a one dimensional Su-Schrieffer-Heeger (SSH) model~\cite{ssh, ssh2, Asboth2016_ssh} with unit cell-wise staggered onsite quasiperiodic disorder we show a remarkable disorder induced charge pumping by periodically modulating the system parameters along a single pumping cycle only. We find that the robust charge pumping in this case is due to an onset of a transition from a topological phase in the clean limit to a trivial phase through a gap closing point as a function of disorder. Surprisingly, the charge pumping in this case can also be achieved by periodically modulating the disorder strength by following a single cycle pumping protocol as opposed to the standard scheme of modulating the hopping strength.

{\em Model.-} 
The system under consideration is a one dimensional lattice of spinless fermions with $N$ two-site unit cells and cell wise staggered onsite quasiperiodic disorder which is written as
\begin{align}
\label{eq:ham_topdis_ssh}
 H=&-t_1\sum_{j=1}^N\left(c_{j,A}^{\dagger}c_{j,B}+H.c.\right)
 \\ \nonumber
    &-t_2\sum_{j=1}^{N-1}\left(c_{j+1,A}^{\dagger}c_{j,B}+H.c.\right)\\ \nonumber
&+\lambda\sum_{j=1}^N (-1)^{j+1} \cos\left[2\pi\beta (2j-1)+\phi\right]n_{j,A}\\ \nonumber
&+\lambda\sum_{j=1}^N (-1)^{j+1} \cos\left[2\pi\beta (2j)+\phi\right]n_{j,B}.
%&+\Delta\sum_{j=1}^N(n_{j, A} - n_{j, B}) \nonumber
\end{align} 
Here, $c_{j,A}$ ($c_{j,B}$) is the fermionic annihilation operator at $A$ ($B$) sublattice site of the $j$th unit cell and $n_{j,A}$ ($n_{j,B}$) is the respective number operator. $t_1$ and $t_2$ are the intra- and inter-cell hopping amplitudes respectively. $\lambda$ sets the strength of the quasiperiodic disorder which we refer to as disorder in the following. The cell-wise staggered nature of the disorder strength produces a pattern like $\lambda, \lambda, -\lambda, -\lambda, \ldots$ throughout the lattice as depicted in Fig.~\ref{fig:figure0}. We consider even number of unit cells so that we have equal number of lattice sites ($L=2N$) with positive and negative disorder strengths in the system. The quasiperiodicity of the lattice is ensured by considering $\beta=(\sqrt5-1)/2$ - the inverse golden mean and all the numerical simulations are performed under periodic boundary conditions, unless otherwise explicitly mentioned. For convenience we fix $t_2=1$ as the energy scale of the system. With this choice, the value of $t_1$ defines the dimerization in the model, i.e., for $t_1\neq 1$ the system is dimerized. In our studies we set $\phi=0$, as the results do not change significantly for finite values of $\phi$ when the system size is sufficiently large.  

In the following we show that the model described above exhibits quantized TCP as a function of disorder. Before moving on to discuss the scenario of charge pumping, we first investigate the topological properties of the model which provides a platform for the TCP.

\begin{figure}[t]
\begin{center}
\includegraphics[width=1\columnwidth]{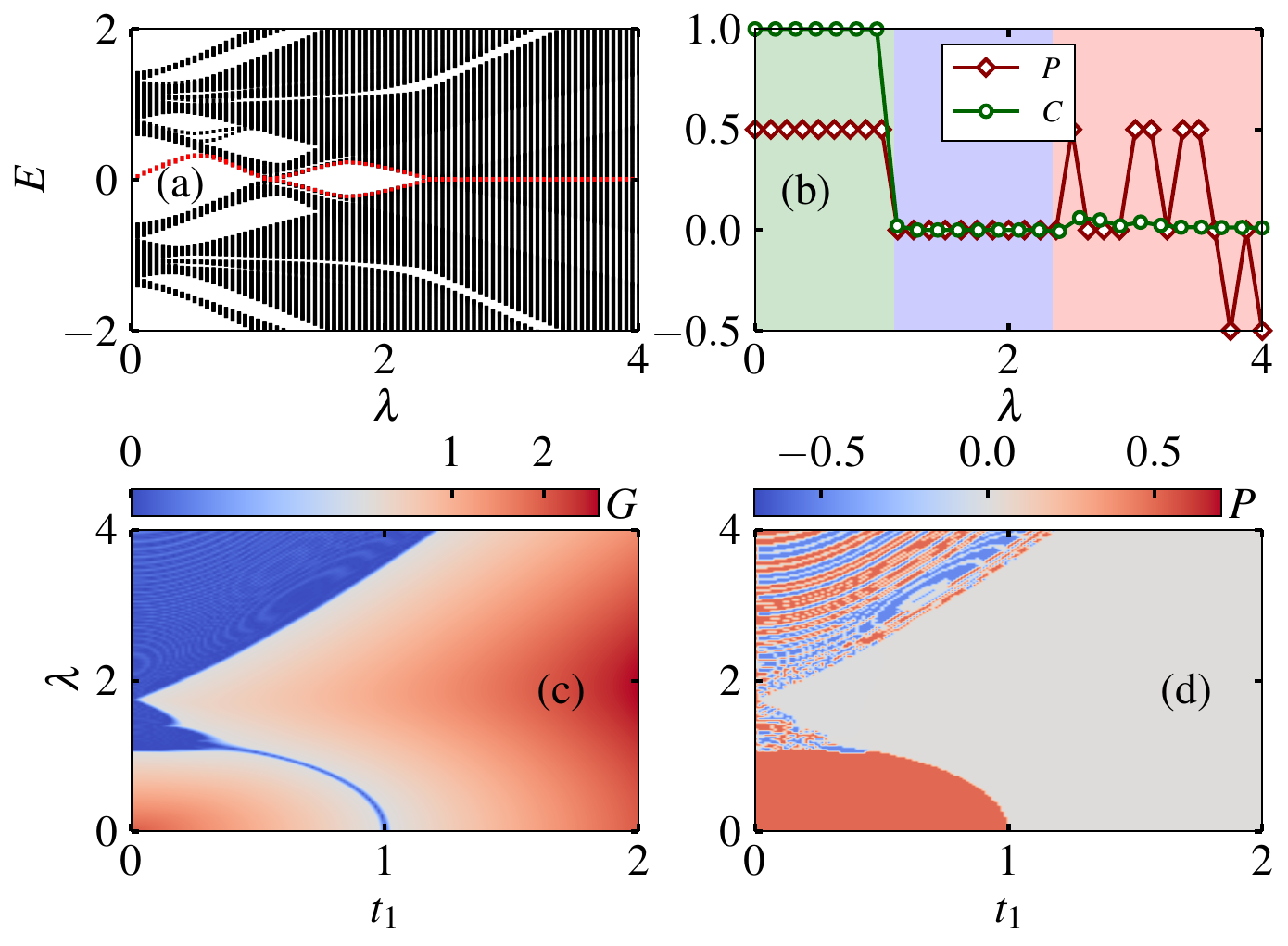}
 \end{center}
\caption{(a) The energy spectrum under open boundary condition for $t_1=0.4$. The pair of red dotted lines highlight the edge states. (b) The polarization $P$ (brown diamonds) and the local Chern marker $C$ (green circles) as a function of $\lambda$ for $t_1=0.4$. The green, blue and red shaded areas mark the topological, trivial and gapless phases, respectively. $G$ and $P$ plotted as a function of $t_1$ and $\lambda$ in (c) and (d) respectively depicting the complete phase diagram of the model. Here we consider $L=2584$ lattice sites. For the calculation of $C$ we consider $J=0.7, \delta_0=0.3$ and $\Delta_0=0.5$ (see text for details).}  
\label{fig:figure1}
\end{figure}

{\em Topological phase transition.-}
As already mentioned in the introduction, topological phases are not stable in the limit of strong disorder and an increase in disorder results in the closing of the bulk gap. However, for the model under consideration we obtain an exception to this behaviour where we find that the gap reopens for a range of dimerization strengths as a function of disorder. In  Fig.~\ref{fig:figure1} (a) we show the single particle energy spectrum as a function of $\lambda$ for $t_1=0.4$ and $L=2584$ imposing open boundary condition. Clearly, when $\lambda=0$, the model shown in Eq.~\ref{eq:ham_topdis_ssh} is the SSH model and for the choice of dimerization, i.e. for $t_1=0.4$, it exhibits a gapped phase which is topological in nature hosting a pair of zero energy edge states (states shown in red in Fig.~\ref{fig:figure1} (a)). However, as $\lambda$ increases, the bulk gap slowly vanishes at $\lambda\sim 1.1$ and then reopens before vanishing again after $\lambda \sim 2.35$ due to strong disorder. On the other hand, the zero energy modes become energetic as $\lambda$ increases and eventually merge with the bulk bands in the first gapped phase. It is important to note that here the edge states are sensitive to the phase $\phi$ and thus the position of them merging into the bulk may vary for different disorder realizations.

To characterise the nature of these gapped phases depicted in Fig.~\ref{fig:figure1}(a), we compute a real space polarization~\cite{Resta_rev, Resta_pol1, pol1, pol2} defined by the formula
\begin{equation}
    P = \frac{1}{2\pi} \Im \ln [\bra{\Psi} e^{\frac{2\pi i}{N}X}  \ket{\Psi}],
\end{equation}
which serves as an invariant to distinguish the topological phase from the trivial and gapless phases. Here, $|\Psi\rangle$ is the many-body ground state at half filling constructed out of the single particle states and $X=\sum_{j=1}^Nj(n_{j, A} + n_{j, B})$ is the total position operator~\cite{Resta_1998}.  In Fig.~\ref{fig:figure1}(b), we plot $P$ as a function of $\lambda$ for $t_1=0.4$ which corresponds to the situation shown in Fig.~\ref{fig:figure1}(a). It can be seen that $P=0.5$ throughout the first gapped phase which is the signature of a topological phase. As $\lambda$ increases, $P$ sharply changes to a value $P=0.0$ at the gap closing point and remains constant in the second gapped phase which is the trivial phase. However, in the gapless region, i.e for $\lambda \gtrsim 2.35$, $P$ does not take any fixed value since the topology is ill-defined there. 

With this information in hand we now try to capture the topological properties of the model as a function of $\lambda$ for different values of dimerization (i.e., $t_1$). To this end we first plot the single particle energy gap between the two middle eigenstates (the bulk gap), $G=E_{L/2+1}-E_{L/2}$ as a function of $t_1$ and $\lambda$ in Fig.~\ref{fig:figure1} (c). As expected, in the absence of  disorder (i.e. along the $t_1$-axis where $\lambda=0$) the gap is finite for all values of $t_1$ except for $t_1=1$ which is the gap-closing critical point of the SSH model. As already well known, the regions for $t_1<1$ and $t_1>1$ are  topological and trivial in nature, respectively. However, as  $\lambda$ is turned on, we obtain that the gap closing critical point shifts towards smaller values of $t_1$ and eventually becomes a gap closing region close to the $\lambda$-axis and for $t_1=0$ (on the $\lambda$-axis) the second gapped region completely vanishes. 

These findings not only highlight a well defined topological phase as a function of $t_1$ for a range of $\lambda$ but also a topological transition as a function of $\lambda$ for a range of $t_1$. 
To quantify the phase diagram we plot the polarization $P$ as a function of $t_1$ and $\lambda$ in Fig.~\ref{fig:figure1} (d). 
The values of $P=0.5$ (brick red region) and $0.0$ (gray region) identify the topological and trivial regions, respectively, whereas in the gapless region $P$ is not well defined (region with no fixed color).

Until now we observe a well defined gap closing transition as a function of onsite quasiperiodic disorder strength. Such a feature is counter-intuitive as strong onsite disorder leads to the overlap of single particle states and thus the emergence of another gapped phase after gap closing is non-trivial. In the following we will discuss the TCP due to such emergent gap-gap transition.

{\em Thouless charge pumping.-}
As mentioned earlier, the TCP enables a quantized transport of particles in a lattice by following a pumping cycle where some of the system parameters are periodically modulated. It is worth noting that a finite and quantized pumping requires three important criteria to be followed during the cycle: (a) the bulk gap should remain open, (b) a well defined topological to trivial phase transition should occur, and (c) the gapless critical point of transition should lie inside the cycle. Traditionally in the SSH model and its variants the TCP is achieved by invoking the celebrated Rice-Mele (RM) model~\cite{rm, Asboth2016_rm} where an external staggered potential keeps the gap open throughout the pumping cycle. In our case, addition of such staggered terms to the model~\ref{eq:ham_topdis_ssh} results in a disordered RM model $H_{\text{dRM}}=H + \Delta\sum_{j=1}^N(n_{j, A} - n_{j, B})$.
Note that in the absence of disorder, i.e. for $\lambda=0$, $H_{\text{dRM}}$ reduces to the original RM model where quantized charge pumping has been observed in different context. 
To implement charge pumping in the presence of disorder,  we first modify the intra-cell and inter-cell hopping amplitudes as $t_1=J-\delta$ and $t_2=J+\delta$, respectively for convenience, where $J$ is a fixed hopping strength and $\delta$ is the dimerization strength. Then we introduce a pumping parameter $\tau$ to periodically modulate the dimerization as $\delta=\delta_0\cos(\tau)$ and staggered potential as  $\Delta=\Delta_0\sin(\tau)$ as depicted in Fig.~\ref{fig:figure0}(b).

\begin{figure}[t]
\begin{center}
\includegraphics[width=1\columnwidth]{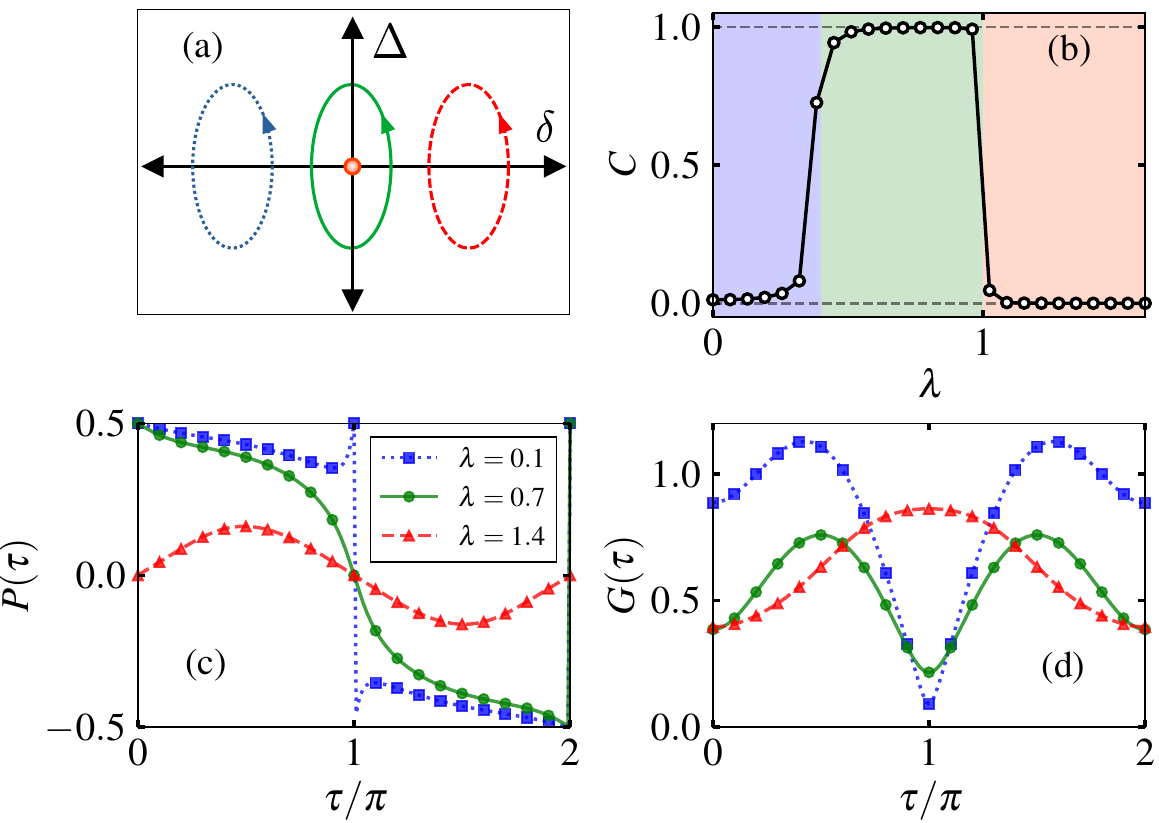}
 \end{center}
\caption{(a) Various pumping cycles in the $\delta-\Delta$ plane. Here the red dot represents the gapless critical point. (b) Local Chern marker $C$ as a function of $\lambda$ for $J=0.75, \delta_0=0.2$ and $ \Delta_0=0.5$. The blue, green and red shaded areas corresponds to the pumping cycles marked by the blue, green and red color cycles respectively in (a). (c) The polarization $P(\tau)$ as a function of $\tau$ for $\lambda=0.1$ (blue squares), $\lambda=0.7$ (green circles) and $\lambda=1.4$ (red triangles). (d) The corresponding bulk gaps $G(\tau)$ are shown as a function of $\tau$. Here we consider $L=2584$ lattice sites.}  
\label{fig:figure2}
\end{figure}

With this setup, we first quantify the topological phase transition as a function of the dimerized hopping for different strengths of disorder. To this aim we compute a quantity known as the local Chern marker (LCM)~\cite{Resta_lcm, hayward} which has been proven to be a suitable topological invariant for inhomogeneous systems such as the one considered here. 
Following the prescription provided in Ref.~\cite{hayward, Resta_lcm}, the redefined LCM in real space position basis is  
\begin{align}
\label{eq:lcm}
    C_j=\frac{1}{\pi} \sum_{n=0}^{N_\tau-1}\Im \bra{j}X_e^{\dagger}\mathcal{P}(\tau_n)X_e \mathcal{Q}(\tau_n) \mathcal{P}(\tau_{n+1})\mathcal{P}(\tau_n)\ket{j},
\end{align}
where $\mathcal{P}$ and $\mathcal{Q}=\mathds{1}-\mathcal{P}$ are the projection operators onto the occupied and unoccupied states of the Hamiltonian $H_{\text{dRM}}$, respectively, and $X_e=\exp(X)$ is the exponentiated position operator. The parameter $\tau$ is discretized with a step size $d\tau=\tau_{n+1}-\tau_n$ and $N_{\tau}$ number of points in the interval $[0,~2\pi]$. The sum $C=\sum_j C_j$ counts the number of pumped charge per cycle and gives rise to the Chern number associated to a phase. In this case $C$ assumes quantized values as long as the bulk gap remains open in the cycle. To capture such behavior we calculate $C$ for each values of $\lambda$ for $J=0.7, \delta_0=0.3$ and $\Delta_0=0.5$ such that after a complete cycle the initial hopping combination at $\tau=0$ ($t_1=0.4, t_2=1$) returns to itself at $\tau=2\pi$ through the point $ t_1=1, t_2=0.4$ when $\tau=\pi$. As anticipated, inside the topological phase we obtain $C=1$ and inside the trivial phase $C=0$  as shown in Fig.~\ref{fig:figure2} (b) (green circles) revealing a quantized TCP. On the other hand, in the gapless phase $C$ is not well defined and does not possess any quantized values.

From this analysis we obtain a clear signature of quantized TCP as the system traverses from the topological to trivial phase transition following the standard pumping protocol for each fixed disorder strength. In the rest of the manuscript, we will show a quantized TCP when disorder is tuned.

{\em Disorder induced Thouless pumping.-} 
As already demonstrated by different studies, disorder (random or quasiperiodic) when strong, is usually considered as an obstruction to robust charge pumping which breaks down due to the closing of the bulk gap. Surprisingly, in the situation under consideration, disorder can actually pump a quantized amount of charge in a lattice which is not possible in the clean limit. To demonstrate this we start with a pumping cycle which is completely within the topological phase of the phase diagram shown in Fig.~\ref{fig:figure1}(c), i.e., the cycle does not connect the trivial phase at all. This implies that the gapless critical point lies outside the cycle when disorder is absent (blue dotted cycle in Fig.~\ref{fig:figure2}(a)). For this purpose, we fix the inter-cell hopping $t_2=1$, and modulate only the intra-cell hopping as $t_1=J-\delta_0\cos(\tau)$ and staggered potential as $\Delta=\Delta_0\sin(\tau)$. We consider the parameter set $(J, \delta_0, \Delta_0) = (0.75, 0.2, 0.5)$, that gives $t_1=0.55$ ($=0.95$) at $\tau=0$ ($=\pi$), which is always less than $t_2$ throughout the pumping cycle. Now, from the computed values of $C$ shown in Fig.~\ref{fig:figure2}(b) we observe that no particle is pumped when $\lambda=0$ and this feature is persistent up to $\lambda\sim 0.5$ (blue shaded region ). However, in the range $0.5\lesssim\lambda\lesssim 1$ (green shaded region ), a remarkable situation happens when we obtain $C=1$ which is an indication of a quantized TCP. The reason behind this quantized TCP is that with increase in $\lambda$, the critical point or the gap closing point for the topological to trivial transition shifts to lower $t_1$ values and gets enclosed by the pumping cycle (the green solid cycle in Fig.~\ref{fig:figure2}(a)). When $\lambda$ increases further, $C$ vanishes again for $\lambda\gtrsim1.0$ (red shaded region ) as the gapless point moves out of the cycle (the red dashed cycle in Fig.~\ref{fig:figure2}(a)). We call this feature of quantized pumping for finite disorder which was absent in the clean limit, as a disorder induced Thouless charge pumping. 

To reconfirm this TCP, we compute the polarization $P(\tau)$ throughout the pumping cycle for all the three cases in Fig.~\ref{fig:figure2}(c) which provides the information about the total pumped charge $Q=\int_{0}^{2\pi}\partial_{\tau}P(\tau)d\tau$. When $\lambda=0.7$, $P(\tau)$ varies smoothly throughout the cycle from $0.5$ to $\sim -0.5$, which indicates a finite and quantized pumping (green circles). However, for $\lambda=0.1$, although the polarization becomes finite, it shows a sharp discontinuity at $\tau=\pi$ (blue squares), indicating no charge being pumped . Similarly, for $\lambda=1.4$ the polarization varies around $P(\tau) = 0$ (red triangles) for all the values of $\tau$ which is a signature of zero charge being pumped . Note that in all the three cycles, the bulk gap of the spectrum remains open throughout as shown in  Fig.~\ref{fig:figure2}(d) where the bulk gap $G(\tau)$ is plotted as a function of $\tau$.  However, only the cycle corresponding to the green curve exhibits a robust pumping as it encloses the gapless critical point (see Fig.~\ref{fig:figure2} (a)).

The above analysis reveals that a robust and quantized Thouless pump can be established at some intermediate values of disorder strength which is initially absent at zero or weak disorder strengths. Such quantized TCP eventually breaks down when the strength of the disorder becomes stronger leading to the closing of the bulk gap. Such non-trivial TCP is possible only due to the emergent topological phase transition as a function of disorder as depicted in Fig.~\ref{fig:figure1}. 

\begin{figure}[!t]
\begin{center}
\includegraphics[width=1\columnwidth]{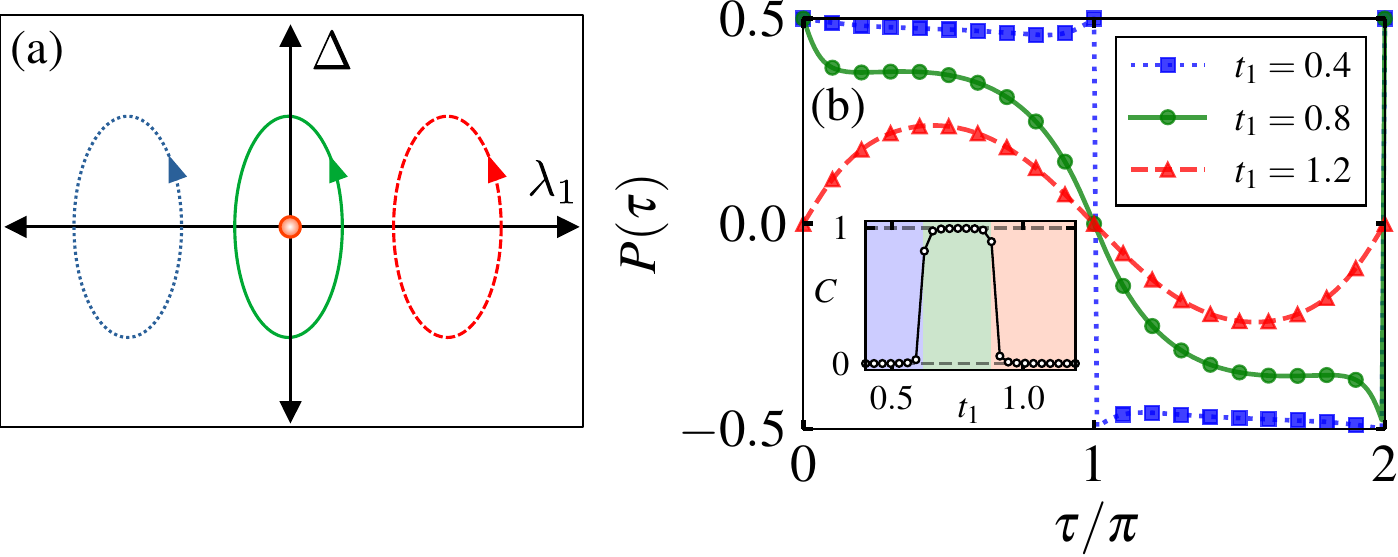}
 \end{center}
\caption{(a) Different pumping cycles in the $\lambda_1-\Delta$ plane. Here the red dot represents the gapless critical point. (b) $P(\tau)$ as a function of $\tau$ for $t_1=0.4$ (blue squares), $t_1=0.8$ (green circles) and $t_1=1.2$ (red triangles). The inset shows local Chern marker $C$ as a function of $t_1$ for $\lambda^{'}=0.75, \lambda_0=0.2$ and $\Delta_0=0.5$. The blue, green and red shaded regions corresponds to the pumping blue, green and red colored pumping cycles respectively in (a). Here we consider $L=2584$ lattice sites.}  
\label{fig:figure3}
\end{figure}

The scenario shown in Fig.~\ref{fig:figure1}(c) also hints at a possibility of a different kind of TCP with respect to the disorder. To establish this scheme we propose a non-standard pumping protocol where the disorder in the $H_{\text{dRM}}$ is periodically modulated in place of the hopping terms (or equivalently dimerization) along with the onsite staggered term $\Delta$. To achieve this we redefine the disorder strength as  $\lambda=\lambda^{'}-\lambda_1$. While we keep  $t_1,~t_2$ and $\lambda^{'}$ fixed, we periodically vary $\lambda_1$ and $\Delta$ as $\lambda_0\cos(\tau)$ and $\Delta_0\sin(\tau)$, respectively. Here three distinct pumping cycles are possible when $t_1$ is increased; Cycle-$1$: the cycle does not connect the trivial phase at all, thus, the gapless critical point lies outside it but the gap is open throughout (blue dotted cycle in Fig.~\ref{fig:figure3}(a)), Cycle-$2$: the cycle encloses the gapless point and the gap also remains open (green solid cycle in Fig.~\ref{fig:figure3}(a)), and Cycle-$3$: the gap remains open but the topological phase is not connected to it (red dashed cycle in Fig.~\ref{fig:figure3}(a)). To investigate the pumping, we plot the polarization $P(\tau)$ as a function of $\tau$ for $t_1=0.4, 0.8$ and $1.2$ which follow the Cycle-1, Cycle-2 and Cycle-3, respectively. We consider the parameter set $(\lambda^{'}, \lambda_0, \Delta_0)=(0.75, 0.2, 0.5)$ that gives $\lambda=0.55$ ($=0.95$) at $\tau=0$ ($=\pi$). When $t_1=0.8$, $P(\tau)$ varies smoothly from $0.5$ to $\sim-0.5$ (green circles) over the cycle indicating a robust pumping. On the other hand, for $t_1=1.2$, there is no pumping as indicated by the polarization varying around $P(\tau)=0$ (red triangles). When $t_1=0.4$, $P(\tau)$ although remains finite in the pumping cycle, a sharp discontinuity is observed near $\tau=\pi$ (blue squares) which results in a zero net charge pump. To further quantify this we compute the local Chern marker $C$ using Eq.~\ref{eq:lcm}. In the inset of Fig.~\ref{fig:figure3}(b), we plot $C$ as a function of $t_1$ for the same parameter set. As expected, when the pumping path follows the Cycle-$1$ for $t_1\lesssim0.6$, $C$ vanishes indicating no pumping of particles (blue shaded area). However, in the regime $0.6\lesssim t_1\lesssim0.9$, we obtain $C=1$ signifying a robust and quantized pumping of particles per cycle (green shaded area). For $t_1 \gtrsim 0.9$, $C$ vanishes again indicating no net charge is pumped (red shaded area). This quantized variation of $C$ according to such a non-standard pumping protocol exhibits a robust TCP in a quasiperiodic lattice.

{\em Conclusions.-} 
In our studies, we have shown that in a one dimensional SSH chain a unit cell-wise staggered onsite quasiperiodic disorder can favor a robust and quantized TCP following the standard protocol of charge pumping which is otherwise unstable and breaks down with increase in disorder strength. In the present case such quantized TCP is achieved due to an emergent topological phase transition from an initial topological phase to another gapped trivial phase through gap closing points for a range of hopping dimerization of the SSH model. By implementing different pumping protocols in the framework of the Rice-Mele model we have shown that disorder can induce a quantized charge pumping. Moreover, we have proposed a completely different pumping protocol where disorder is directly involved in the charge transport mechanism. 

Our finding not only addresses an unanswered question on the fate of quantized TCP in the presence of disorder but also allows quantised charge pump following the standard pumping protocols which was not shown earlier in any studies involving onsite disorder. Given the recent interest in the study of quantized charge pumping in topological systems and recent experimental observations of TCP, our findings will provide impetus to open up a new direction in the field. An immediate extension can be to search for other models where such disorder driven charge pumping can be obtained. The question can be asked if both the hopping and onsite quasiperiodic disorder are turned on. The extension to two dimension and the effect of interaction in such systems can also be explored.

{\em Acknowledgment.-}
T.M. acknowledges support from Science and Engineering Research Board (SERB), Govt. of India, through project No. MTR/2022/000382 and STR/2022/000023.

\bibliography{references}

\end{document}